\documentstyle[sprocl]{article}
\input{epsf}
\def\be{\begin{equation}}
\def\ee{\end{equation}}
\def\bea{\begin{eqnarray}}
\def\eea{\end{eqnarray}}

\def\tr{{\rm Tr}}

\def\Hol{{\rm Hol}}
\def\tr{{\rm Tr}}

\def\bea{\begin{eqnarray}}
\def\eea{\end{eqnarray}}

\def\bb#1{\hbox{\mybb#1}}

\def\R4{\bb{R}^4}
\font\mybb=msbm10 at 12pt

\def\tr{{\rm Tr}}

\def\Hol{{\rm Hol}}
\def\tr{{\rm Tr}}

\def\unita{{1 \kern-.30em 1}}

\begin{document}

\title{THE BF FORMALISM FOR YANG-MILLS THEORY AND THE `t HOOFT ALGEBRA}

\author{ MAURIZIO MARTELLINI$^\dagger$ and MAURO ZENI }

\address{Dipartimento di Fisica, Universit\`a di Milano and 
I.N.F.N.\ - \ Sezione di Milano.\\ 
$\dagger$ Landau Network  at ``Centro Volta'', Como, ITALY}


\maketitle\abstracts{  
The deformation of a topological field theory, namely the pure BF theory, 
gives the first order formulation of Yang-Mills theory; Feynman rules are 
given and the standard {\it uv}-behaviour is recovered. 
In this formulation new non local observables can be introduced following the 
topological theory and giving an explicit realization of `t Hooft algebra.
}
\section{The BF formulation of Yang-Mills theory}
We consider the following euclidean first order action 
functional \cite{fmz,ccgm,hr}
\be 
S_{BFYM} = \int_{M^4} d^4x {i\over 2}\varepsilon^{\mu\nu\alpha\beta}
B^a_{\mu\nu}F^a_{\alpha\beta}
+g^2 \int_{M^4} d^4x B^a_{\mu\nu}B^{a\mu\nu} \quad ,
\label{uno.1}
\ee  
where
$F^a_{\mu\nu}\equiv 2\partial_{[\mu}A^a_{\nu ]} +f^{abc}A^b_\mu A^c_\nu $ 
is the field strenght of the gauge field and where $B$ is a Lie valued 2-form; 
the field equations of (\ref{uno.1}) are 
\be
*F^a_{\mu\nu}=2ig^2B^a_{\mu\nu}\qquad ,\qquad 
\varepsilon ^{\mu\nu\alpha\beta}D_\nu B^a_{\alpha\beta} =0\quad ,
\label{uno.2}
\ee
where 
$*F_{\mu\nu}\equiv\frac 12 \varepsilon_{\mu\nu\alpha\beta} 
F^{\alpha\beta}$ 
is the ``dual ''field strength.
This action enjoys the usual gauge invariance 
$\delta A_\mu = D_\mu c$, $\delta B_{\mu\nu}= i[c,B_{\mu\nu}]$ 
with $D_\mu\equiv\partial_\mu -i[A_\mu ,\cdot ]$ and the 
standard Yang-Mills (YM) action is recovered performing path integration over B 
or by substituting the field equations in it.

The first term in the r.h.s. of (\ref{uno.1}) is the action of a 
topological quantum field theory, the so-called BF theory \cite{horouno,blau}. 
The pure BF theory has an extra ``topological'' symmetry 
\be 
\delta A^a_\mu = 0 \qquad ,\qquad 
\delta B^a_{\mu\nu} = 
 2\partial_{[\mu}\psi^a_{\nu ]}  +2f^{abc}A^b_{[\mu} \psi^c_{\nu ]}\quad ,
\label{uno.3}
\ee 
where  $\psi $ is a 1-form. Note that due to zero modes 
$\delta \psi_\mu =D_\mu\phi $ with $\delta\phi =0$ in 
the transformations (\ref{uno.3}) the symmetry is reducible, 
allowing for a ghosts of ghosts 
structure. The presence of this 
topological symmetry cancels out any local dynamics from the 
pure BF theory. In our case local degrees od freedom are restored by 
the second term in r.h.s. of (\ref{uno.1}) which is an explicit 
symmetry breaking for the topological symmetry (\ref{uno.3}) as long as 
$g\neq 0$. Therefore in this formulation YM theory appears as 
a deformation of the topological field theory and we call it BFYM. 

The question arises whether the first order formulation is equivalent to the 
standard one at the quantum  level and to which extent this equivalence holds. 
In particular note that ``on-shell''  $B$ coincides with 
the dual field strength and satisfies the Bianchi identities. This is no longer 
true off-shell and this fact has been related to the presence of 
``monopoles charges''  in the vacuum \cite{fmz} which should enter 
the non perturbative sector of the theory. Moreover , as shall be 
discussed later, in BFYM theory  can be defined new non local observables  
related to the phase structure of the theory. 
Indeed from the point of view of the perturbative regime we expect 
the same $uv$-behaviour as in standard YM. 

\section{Perturbative formulation}
~When casting a perturbative framework in BFYM we actually have two different 
ways of quantizing the action (\ref{uno.1}). 
The first one is to include the 
topological symmetry breaking term $g^2B^2$ in the kinetic operator \cite{mz}, 
the 
second one is to regard it as a vertex. The latter procedure is quite involved 
because the kinetic operator in this case requires gauge fixing also for the 
topological symmetry and a ghosts of ghost structure; moreover the degrees of 
freedom of the topological group become dynamical and add to the field content 
of the theory. We will address to this case elsewhere \cite{fmtz}
and consider now the former one. The gauge fixing is straigthforward and, 
after the field rescaling $B\to B/g$ and $A\to gA$, the 
following Feynman rules are easily obtained:
\vskip .3cm
\centerline{\vbox{\epsfysize=6.5cm\epsfbox{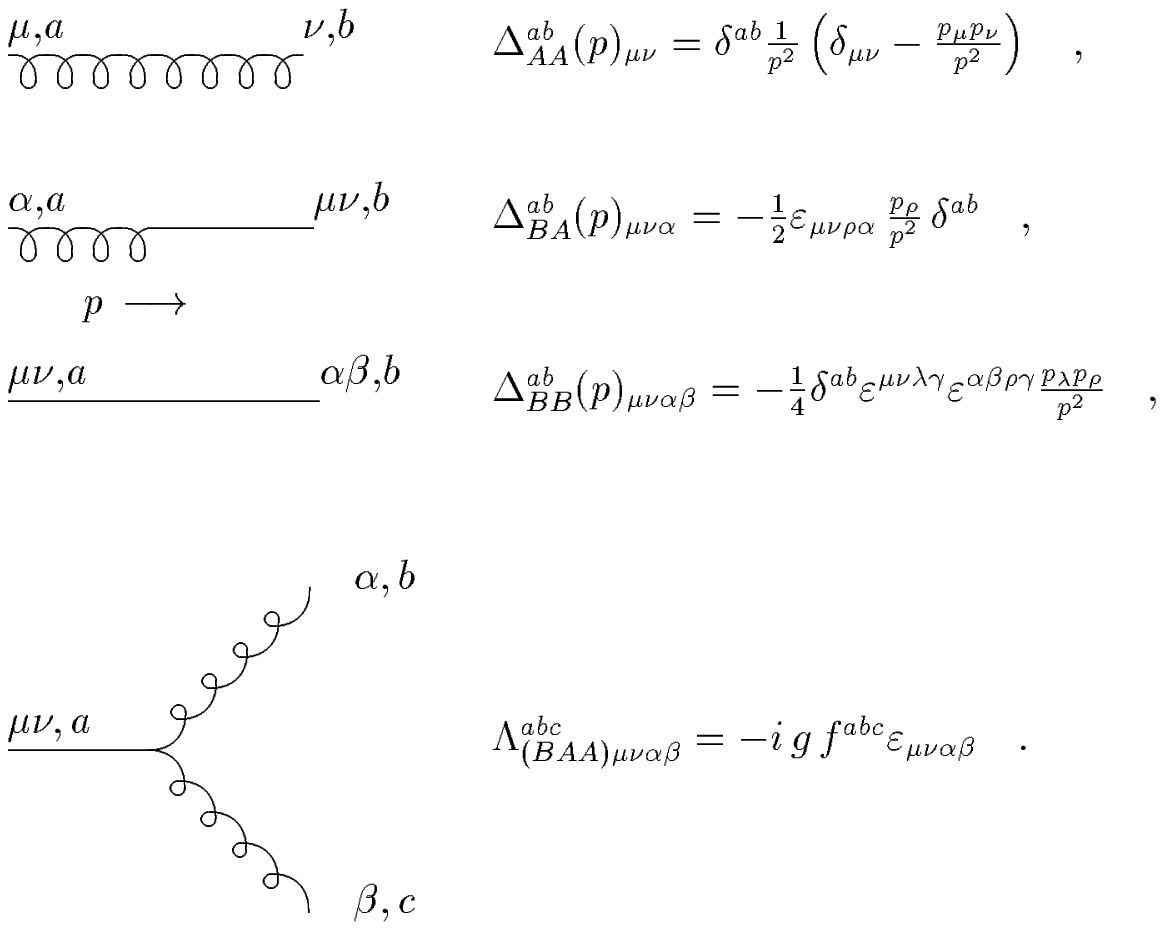}}}
%
Note the off-diagonal structure of the propagators. Indeed is this structure 
which recovers all the nonlinear interactions of YM, even if our theory has 
only the trilinear vertex $BAA$. 
The ghost sector is unchanged in this formulation and its Feynman rules have 
to be added to the previous ones.

When studying one loop diagrams one finds in particular that the singular 
structure of the propagator $\Delta_{AA}$ is given in the Landau gauge by 
\vskip .3cm
\centerline{\vbox{\epsfysize=3cm\epsfbox{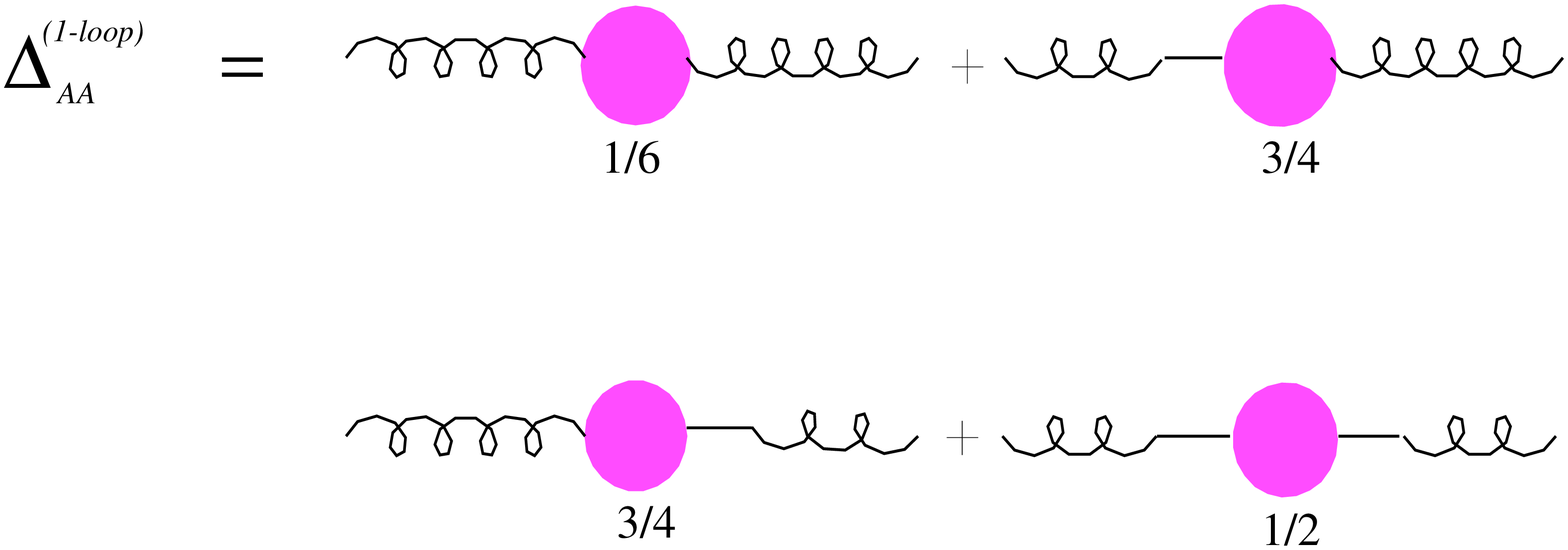}}}
\noindent 
and requires for $A$ the same wave function renormalization found in YM, i.e.
$A_0=Z_{AA}A_R$ with $Z_{AA}$ given in dimensional regularization by  
$Z_{AA}=1 +\frac{13}{12}\frac{c_Vg^2}{16\pi^2\varepsilon}$. This value 
leads to the expected one loop $\beta$-function, 
$\beta_1=\frac{11}3\frac{c_V}{16\pi^2}$, 
and therefore the 
$uv$-behaviour of YM and BFYM are the same \cite{mz}.

When performing the renormalization of the theory the dimensions and tensorial 
structure of the fields allow the operator mixing
\be 
\left( \begin{array}{c} B_0\\F_0 \end{array}\right) =
\pmatrix{Z_{BB} & i*Z_{BA}\cr 0&Z_{AA} }\cdot 
\left( \begin{array}{c} B_R\\F_R \end{array}\right)
\quad .
\ee
In this way is generated a counterterm $\sim F^2$, required to renormalize the 
non linear gluon vertices which arise at the one loop level. Note that 
$Z_{AB}\sim g^2\frac 1{\varepsilon}$ in order to not modify Feynman rules at 
the tree level. After renormalization is performed is always possible to 
redefine $B_R$ in order to reabsorb the $F^2$ term and recover the tree level 
structure of the theory.

\section{A new observable}
A new non-local observable associated to an orientable surface 
$\Sigma\in M^4$  is naturally introduced in the BF formulation 
of QCD \cite{fmz,ccgm},
\be                       
M(\Sigma ,C)\equiv \tr \exp \{ik
\int_{\Sigma } d^2y 
\ \Hol^y_{\bar x} (\gamma ) B(y) \Hol_y^{\bar x} (\gamma^{\prime} )\} \quad ,
 \label{due.1} 
\ee 
where $\Hol_{\bar x}^y(\gamma )$ denotes the  holonomy 
along the open path $\gamma\equiv\gamma_{\bar x y}$ with initial and final 
point $\bar x$ and $y$ respectively,
\be 
\Hol_{\bar x}^y(\gamma )\equiv P\exp \{i\int_{\bar x}^y dx^{\mu} A_{\mu}(x)\}
\quad .
\label{due.2}
\ee 
In (\ref{due.1}) $k$ is an arbitrary parameter, 
$\bar x$ is a {\it fixed} point 
and the relation between the assigned paths 
$\gamma$, $\gamma^{\prime}$ over $\Sigma$ and the closed contour $C$ is the 
following: $C$ 
starts from the fixed point $\bar x$, connects a point $y\in\Sigma$ by 
the open path $\gamma_{\bar x y}$ and then returns back to the
neighborhood of $\bar x$ by 
$\gamma_{y\bar x }^{\prime}$, 
(in general $(\gamma_{\bar x y})^{-1}\neq\gamma_{y\bar x }^{\prime}$). 
From the neighborhood 
of $\bar x$ the path starts again to connect another point 
$y^{\prime}\in\Sigma$. Then it 
returns back 
and so on until all the points
on $\Sigma$ are connected. The path $C$ is generic and 
no special ordering prescription is required. Of course the 
quantity (\ref{due.1}) is path dependent and our strategy is to regard it as 
a loop variable once the surface $\Sigma$ is given. 

In particular it is possible to show \cite{fmz} that $M$ 
gives an explicit analytic 
realization of the `t Hooft loop operator \cite{thooft1}, the dual  to the 
Wilson loop $W$. The $vev$'s of $M$ and $W$ label the phases of the theory and 
an explicit calculation in the BF formalism 
has shown the expected confining phase. Lastly, a quite similar formalism has 
been introduced recently in \cite{poly} for the $vev$ of $W$  
in a string picture of gauge theories.

This work has been partially supported by MURST and by 
TMR programme  ERB-4061-PL-95-0789 in 
which M.~Z. is associated to Milan.

\end{document}